\documentclass[journal]{IEEEtran}
\usepackage{ifpdf}
\usepackage{cite}
\usepackage{algorithmic}
\usepackage[ruled,vlined]{algorithm2e}
\ifCLASSINFOpdf
   \usepackage[pdftex]{graphicx}

\else
 \usepackage[dvips]{graphicx}
\fi
\usepackage[cmex10]{amsmath}
\usepackage{epstopdf}
\usepackage{array}
\usepackage{flushend}
\usepackage{balance}
\usepackage{color}

\usepackage{eqparbox}

\usepackage[tight,footnotesize]{subfigure}

\usepackage{fixltx2e}

\usepackage{float}

\usepackage{dblfloatfix}
\usepackage{url}
\usepackage{cite}
\usepackage{amsthm}
\newcommand*{\Scale}[2][4]{\scalebox{#1}{$#2$}}

\usepackage{amssymb}
\begin{document}
\title{Physical Layer Security in Cooperative NOMA Hybrid VLC/RF Systems}

\setlength{\columnsep}{0.21 in}



\author{Mohanad~Obeed,~ \thanks{ M. Obeed, and A. Chaaban are with the School of Engineering, University of British Columbia (UBC), Kelowna, British Columbia, Canada (email:  mohanad.obeed@ubc.ca, anas.chaaban@ubc.ca).}
        Anas Chaaban,~\IEEEmembership{Member,~IEEE,}
        
        Anas~M.~Salhab,~\IEEEmembership{Senior Member,~IEEE \thanks{ A. M. Salhab, and S. A. Zummo are with the Department Electrical Engineering, King Fahd University of Petroleum \& Minerals (KFUPM), Dhahran, Eastern Province, Saudi Arabia (email:  salhab@kfupm.edu.sa, zummo@kfupm.edu.sa).}}
        Salam~A.~Zummo,~\IEEEmembership{Senior Member,~IEEE,} and
        Mohamed-Slim~Alouini,~\IEEEmembership{Fellow,~IEEE} \thanks{M.-S. Alouini is with the Department Computer, Electrical, and Mathematical Siences \& Engineering, King Abdullah University of Science and Technology (KAUST), Thuwal, Makkah Province, Saudi Arabia (email: slim.alouini@kaust.edu.sa).}}
\maketitle
\begin{abstract}
Integrating visible light communication (VLC) and radio-frequency (RF) networks can improve the performance of communication systems in terms of coverage and data rates. However, adding RF links to VLC networks weakens the secrecy performance due to the broadcast and ubiquitous nature of RF links.  This paper studies the physical layer security (PLS) in cooperative non-orthogonal multiple access (CoNOMA) hybrid VLC/RF systems. Consider a VLC system, where two entrusted users close to a VLC access point (AP) help an out-of-coverage legitimate user using RF signals in the presence of an eavesdropper. The AP transmits data to both entrusted users and the legitimate user using the principle of NOMA, where the entrusted users harvest energy from the received light intensity, decode the legitimate user's message, forward it using a RF link, and then decode their messages. It is required to maximize the secrecy rate at the legitimate user under quality-of-service (QoS) constraints using beamforming and DC-bias and power allocation. Different solutions are proposed for both active and passive eavesdropper cases, using semidefinite relaxation, zero-forcing, beamforming, and jamming. Numerical results compare between the different proposed approaches and show how the proposed approaches contribute in improving the secrecy performance of the proposed model.   
\end{abstract}
\IEEEpeerreviewmaketitle
\section{Introduction}

Due to the increasing need for high data rates and the overcrowded radio-frequency (RF) spectrum, researchers and engineers have recently explored different untapped spectrum to transmit data. To exploit the wide license-free visible light spectrum, visible light communication (VLC) has emerged as a promising technology to supplement RF wireless networks \cite{dang2020should}. Some previous works on VLC proved that VLC systems is able to provide data rates up to several Giga-bits/second \cite{schrenk2018visible, tsonev2015towards}, which enables them to be qualified to meet the required high data rates in future wireless networks. 

In order to efficiently utilize the available spectrum, non-orthogonal multiple access (NOMA) has been introduced to increase the spectral efficiency and the fairness of the communication systems. The principle of NOMA  is to send messages to multiple users using the same frequency/time resources with different power levels, and use successive interference cancellation at the users to decode the messages.  NOMA has been investigated, evaluated, and optimized in VLC networks \cite{ yin2016performance,marshoud2016non,zhang2017user,chaaban2016capacity}, where it has been shown that NOMA outperforms orthogonal multiple access (OMA) schemes in terms of data rates and fairness.

Applying NOMA in VLC systems doesn't overcome some VLC drawbacks such as inter-cell interference, limited coverage, and blockage. One method to mitigate such drawbacks and improve performance is to utilize cooperation among users using RF links. This cooperation exploits the received strong signal at some users to help other users, which receives weak signals by acting as cooperative relays. This scheme, known as cooperative NOMA (Co-NOMA) was investigated in RF networks where the strong user serves as a decode-and-forward (DF) relay, by  forwarding the weak user's message after decoding it, so that the weak user receives two versions of the signal, which increases the received signal-to-noise (SNR). In VLC systems, it is not practical to relay the VLC received signal at the the strong user using another VLC link. However, RF links can be used to forward the signals instead. Thus, the weak user has two options: either to be served directly by the VLC AP, or to be served through the hybrid VLC/RF link with the help of the strong user. This scheme is helpful in VLC systems to overcome the limited coverage and susceptibilities to blockage of VLC. 
Authors of \cite{CONOMA} showed that Co-NOMA can improve the sum-rate and the fairness of a VLC system that consists of one AP and multiple users and proposed solutions for joint power allocation, link selection, and user pairing. Authors of \cite{CONOMA2} showed that Co-NOMA in VLC networks can mitigate the impact of the inter-cell interference in multi-cell multi-user VLC systems. 

Unfortunately, this form of Co-NOMA in VLC systems introduces another challenge which is security. VLC systems are generally more secure than RF systems due to that VLC links are blocked by objects and can be directed to cover only small areas that contain authorized users. However, using Co-NOMA with RF links to reach out-of-coverage users can compromise security since an eavesdropper may be able to attain confidential information intended to the legitimate user. Thus, it is important to study the security of such a Co-NOMA VLC/RF system. 

Several papers in the literature investigated physical layer security (PLS) in VLC networks \cite{PLS_Survey, mostafa2016optimal, yin2018physical, arfaouiChaaban, ChaabanLamp}, and in hybrid VLC/RF networks \cite{marzban2017beamforming,pan2017secrecy}. Authors of \cite{PLS_Survey} reviewed the work conducted on optimizing the PLS in VLC and free-space optical communication networks.
Authors of \cite{marzban2017beamforming} investigated PLS based on the assumption that the receivers can gather information from VLC and RF APs at the same time.  The authors designed the RF and VLC beamforming vectors to null the information rate at the eavesdropper and to minimize transmit power for energy efficiency purposes. In \cite{pan2017secrecy}, the authors used VLC links for the downlink and RF links for the uplink, and derived the  secrecy outage probability for the RF links when the users use the energy harvested through the received light intensity to forward their signals to the AP. However, to our best knowledge, the security of Co-NOMA hybrid VLC/RF system has not been investigated in the literature. 

Thus, in this work, we investigate and optimize PLS in Co-NOMA hybrid VLC/RF systems. We consider a system model that consists of a single VLC AP, two entrusted users, one destination (legitimate user), and one eavesdropper. The entrusted users are the users that are served by the VLC AP directly, and are trusted to decode and forward the destination's messages. We assume that the destination and the eavesdropper are out-of-the coverage of the VLC AP or blocked (i.e., either far from the AP or in an other room). The goal of the AP is to transmit data to the entrusted users with the required quality and to the destination in a secure way. We assume that transmission to the entrusted users is secured by the VLC coverage, since the eavesdropper can not tap into VLC signal. Thus, it remains to secure the destination user who is served by RF.  

We consider two cases for the eavesdropper: Either the channel-state-information (CSI) of the eavesdropper is available at the transmitters (i.e., the eavesdropper is active in the network, but it is not authorized to access some confidential information), or the CSI is not available (i.e., the transmitters are unaware of the presence of an eavesdropper). When the eavesdropper's CSI is known,  we formulate the problem as maximizing the secrecy capacity at the legitimate user under quality-of-service (QoS) constraints for the entrusted and legitimate users. We propose two solutions for such non-convex problem: The first is based on semidefinite relaxation (SDR) and Charnes-Cooper methods, and the second is based on nulling the signal at the eavesdropper. When the eavesdropper's CSI is unknown, we adopt three approaches to improve the secrecy performance, one is the beamforming approach and the others are based on injecting artificial noise to confuse the eavesdropper (jamming). In all the proposed approaches, we find solutions for the DC-bias, the transmit power, and the beamforming vector at the entrused users.  Numerical results show that when the eavesdropper's CSI is known, the proposed SDR with  Charnes-Cooper method performs better than zero-forcing approach if the eavesdropper is away from the midpoint between of the entrusted users, while zero-forcing is better if the eavesdropper is nearer to the transmitters. If the eavesdropper's CSI is unavailable, then sending artificial noise with beamformers chosen using SDR or maximum ratio transmission (MRT) performs better than beamforming approach, where the weights are designed to match the channel of the destination without emitting jamming signal. Numerical results also show that in the presence or the absence of CSI, the required QoS at the entrusted users significantly compromises the secrecy performance. 


\section{System Model and Problem Statement}

\begin{figure}[!t]
\centering
\includegraphics[width=3in]{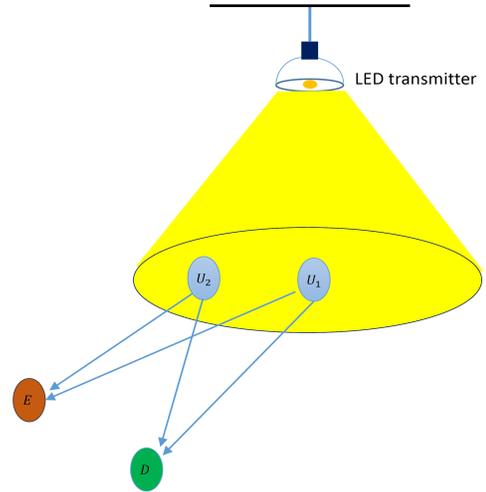}
\caption{A VLC/RF system where the VLC AP transmits to two users within its coverage, which in turn relay information to a destination using an RF link while securing the information from the eavesdropper.}
\label{SM}
\end{figure}

As shown in Fig. \ref{SM}, the considered system consists of one VLC AP, three legitimate users, and one eavesdropper. Cooperative NOMA based on hybrid VLC/RF is used. We assume that two entrusted users ($U_1$ and $U_2$) are in the coverage range of the VLC AP, while the third user $D$ and the eavesdropper $E$ are not. Using cooperative NOMA, the AP sends the three signals of the legitimate users using the same VLC channel, $U_1$ and $U_2$ decode their desired signals in addition to $D$'s signal,  and forward the latter using an RF link. $U_1$ and $U_2$ also harvest energy from the received VLC signal, and use it to forward $D$'s signal. The eavesdropper also receives the forwarded RF signal, which is to be secured.

\subsection{VLC and RF Channel Models}
 The VLC channel between the AP and the user $i$ is given by \cite{wir}
\begin{equation}
\label{vlcch}
h_{i}=\frac{(m+1)A_{p}}{2\pi d_{i}^2} \cos^m(\phi)g_{of}f(\theta)\cos(\theta),
\end{equation}%
where $A_p$ is the photo-detector (PD) physical area, $m=-\left(\log_2(\cos(\theta_{\frac{1}{2}})\right)^{-1}$ is the Lambertian index, $\theta_{\frac{1}{2}}$ is the semi-angle of half power,  $d_{j}$ is the distance between the  AP and the user $i$, $g_{of}$ is  the gain of the optical filter, $\phi$ is the LED  radiance angle, $\theta$ is the PD incidence angle, and $f(\theta)$  is the gain of the optical concentrator given by
\begin{equation}
\label{FoVE}
f(\theta)= \begin{cases}\frac{n^2}{\sin^2(\Theta)}, & \theta\leq\Theta; \\
0, & \theta>\Theta,
\end{cases}
\end{equation}
where $n$ is the refractive index and $\Theta$ is the semi-angle
of the user's field-of-view (FoV).
On the other hand, the RF channel in an indoor environment between users $i$ and $D$ is given by \cite{Perahia2013}
\begin{equation}
h_{D,i}= H^{(RF)}_{D,i} 10^{-\frac{L(d_{D,i})}{20}},
\end{equation}
where $H^{(RF)}_{D,i}$ is  the RF multipath propagation channel, $L(d_{D,i})$ is the path loss, and $d_{D,i}$ is the distance between users $D$ and $i$.
  
\subsection{Transmission Scheme}

Using  NOMA, the AP transmits messages to users $U_1$, $U_2$, and $D$, with rates $R_{1}, R_{2}$, and $R_D$ using the same VLC channel by sending a weighted-sum of the codeword symbols with different weights corresponding to power levels. Since VLC is realized by modulating light intensity, the transmit signal must be nonnegative. Hence, the transmitted optical signal from the AP can be expressed as 
\begin{equation}
y=\nu (\sqrt{P_1}s_1+\sqrt{P_2}s_2+\sqrt{P_D}s_D)+\nu b,
\end{equation}
where $\nu$ is the electrical-to-optical conversion factor measured in W/A, $s_1$, $s_2$, and $s_D$ are codeword symbols assumed to be with unit peak amplitude $\vert s_i \vert<1$, and  $P_1$, $P_2$, and $P_D$ are the peak powers assigned to these symbols, respectively, and $b$ is a DC-bias added to ensure a nonnegative transmit signal. Note that $s_D$ should be forwarded to the destination $D$ by $U_1$ and $U_2$. The messages $s_1$, $s_2$, and $s_D$ are assumed to be with unit power.   
The received electrical signal at the $i$th entrusted user is given by 
\begin{equation}
x_i=\nu \rho h_i(\sqrt{P_1}s_1+\sqrt{P_2}s_2+\sqrt{P_D}s_D)+\nu \rho h_i b+n_i, \ i=1,2
\end{equation}
where $\rho$ is the optical-to-electrical conversion factor measured in A/W,  $h_i$, $i=1,2$ are  channels between the VLC AP and users $U_i$, $i=1,2$,  $n_i$ is a real-valued zero-mean additive white Gaussian noise (AWGN) with variance $\sigma^2_v = N_vB_v$, where $N_v$ is the noise power spectral density, and $B_v$ is the modulation bandwidth. 


Following the NOMA technique, the entrusted users first decode the message of the far user $D$, then they decode their messages afterwards. This imposes the following rate constraints at high SNR, assuming $h_1>h_2$

\begin{align}
\label{Ru1}
R_1\leq R_{u_1}&=\frac{1}{2}\log_2\left(1+\frac{c\rho^2\nu^2h_1^2P_1}{\sigma_v^2}\right),\\
\label{Ru2}
R_2\leq R_{u_2}&=\frac{1}{2}\log_2\left(1+\frac{c\rho^2\nu^2h_2^2P_2}{\sigma_v^2+c\rho^2\nu^2h_2^2P_1}\right),\\
\label{RuD1}
  R_D\leq R_{u_1\rightarrow D}&=\Scale[0.99]{\frac{1}{2}\log_2\left(1+\frac{c\rho^2\nu^2h_1^2P_D}{\sigma_v^2+c\rho^2\nu^2h_1^2P_1+c\rho^2\nu^2h_1^2P_2}\right)},\\
\label{RuD2}
   R_D\leq R_{u_2\rightarrow D}&=\Scale[0.99]{\frac{1}{2}\log_2\left(1+\frac{c\rho^2\nu^2h_2^2P_D}{\sigma_v^2+c\rho^2\nu^2h_2^2P_1+c\rho^2\nu^2h_2^2P_2}\right)},
\end{align}
 where $c=\min\{\frac{1}{2\pi e} \frac{e b^2}{I_H^22\pi}\}$ \cite{chaaban2016capacity} and $e$ is  Euler's number. Since $b$ is restricted to be greater than or equal to $\frac{I_H}{2}$, we have that $c=\frac{1}{2\pi e}$.  

Then, each user forwards the message of user $D$ after encoding it using a secrecy code \cite{el2011network}. In particular, the $i$th user sends $w_i \tilde{s}_D$, where $\vert w_i \vert^2\leq P_{r,i}$ and $\tilde{s}_D$ is the encoded message of user D.
Therefore, the received signal at user $D$ is given by
\begin{equation}
\label{Xd}
x_D=h_{D,1}w_1\tilde{s}_D+h_{D,2}w_2\tilde{s}_D+n_{rf},
\end{equation}
where $h_{D,i}$ is the RF channel between $U_i$ and $D$ and $n_{rf}$ is an additive Gaussian noise with zero mean and variance $\sigma^2_{RF}$.
The received signal at the eavesdropper $E$ is given by
\begin{equation}
\label{Xe}
x_E=h_{E,1}w_1\tilde{s}_D+h_{E,2}w_2\tilde{s}_D+n_{rf},
\end{equation}
where $h_{E,i}$ is the RF channels between $U_i$ and $E$,  and $n_{rf}$ is Gaussian noise with zero mean and  variance $\sigma^2_{RF}$.
We assume that the bandwidth of the RF channel
is a fraction $\eta$ of the bandwidth of the VLC channel, where
$\eta\in(0,1]$. The achievable rate $R_D$ must be smaller than the achievable secrecy rate of the wiretap channel defined by (\ref{Xd}) and (\ref{Xe}), leading to \cite{5961840, 5352243}
\begin{multline}
\label{RD}
\frac{R_D}{\eta}\leq R_s=\frac{1}{2}\log_2\left(1+\frac{\mathbf{h}_D^H\mathbf{w}\mathbf{w}^H\mathbf{h}_D}{\sigma^2_{RF}}\right)\\
-\frac{1}{2}\log_2\left(1+\frac{\mathbf{h}_E^H\mathbf{w}\mathbf{w}^H\mathbf{h}_E}{\sigma^2_{RF}}\right),
\end{multline} 
where $\mathbf{h}_D=[h_{D,1}\  h_{D,2}]^T$, $\mathbf{h}_E=[h_{E,1}\  h_{E,2}]^T$, and $\mathbf{w}=[w_1\ w_2]^T$. 
\subsection{Energy Harvesting}
The entrusted users work also as relays and have the capability to harvest energy from the received VLC signal and use it to relay the $D$'s symbol. 
To transfer the power, a capacitor can separate the DC component from the received electrical signal  and forward it to an energy harvesting circuit \cite{wang2015design, obeed2019dc}. The harvested power at the $i$th user is given by \cite{solar}
\begin{equation}
P_{r,i}=fI_{DC,i}V_{oc,i},
\end{equation}
where $f$ is the fill factor (typically around $0.75$),  $I_{DC,i}=\rho \nu h_ib$ is the DC current received at user $i$, and
$V_{oc,i}= V_t \ln(1+\frac{I_{DC,i}}{I_0}),$ where $V_t$ is the thermal voltage and $I_0$ is the dark saturation current of the PD. Therefore, the harvested electrical power at the $i$th user, which is a function of $b$, can be written as
\begin{equation}
\label{EH}
P_{r,i}(b)=f \rho \nu V_t h_ib\ln\bigg(1+\frac{\rho h_i\nu b}{I_0}\bigg).
\end{equation}

Next, we describe methods to optimize the performance of the system under two considerations: Known and unknown eavesdropper CSI.

\subsection{Problem Statement}
Communication in this system has to satisfy the following constraints. The entrusted users' rates must satisfy a QoS constraint given by $R_i\geq R_{th}$, $i=1,2$. Moreover, the power constraint at the entrusted users must be satisfied, which leads to $|w_i|^2\leq P_{r,i}(b)$ where $P_{r,i}(b)$ is the available power at the entrusted user. This power is related to the amount of harvested optical energy, which depends on the DC bias $b$. To minimize signal clipping and guarantee a postive signal, we require that $b\in[I_H/2,I_H]$. 

Under these constraints, the goal is to maximize the rate $R_D$ under which secrecy can be guaranteed. This problem can be formulated in different ways, depending on the availability of the eavesdropper CSI at the transmitters, or its absence. This problem is discussed in the following sections.
\section{Performance Optimization given Eavesdropper's CSI}
\label{CSI}
In this case, we assume that the CSI of the eavesdropper is known at the AP and at the entrusted users. This allows the AP and the entrusted users to encode D's message at the appropriate rate, so that the message can be sent from the AP to $U_1$ and $U_2$ to $D$, securely.  Hence, our goal is to maximize the secrecy rate of the destination $D$ while satisfying the constraints. 
 The optimization problem can be formulated as follows
\begin{subequations}
\label{RM1}
\begin{eqnarray}
&\displaystyle\max_{\mathbf{w}, b, P_1, P_2, P_D}&  R_s\\
\label{RMb}
&\text{s.t.}&  \vert w_i \vert^2 \leq P_{r,i}(b), \ i=1,2 \\
\label{RMc}
&&  R_{u_i}\geq R_{th},\ i=1,2 ,\\
\label{RMd}
&& R_{u_i\rightarrow D}\geq \eta R_{s},\ i=1,2 ,\\
\label{RMf}
&& \sqrt{P_1}+\sqrt{P_2}+\sqrt{P_D} \leq I_H-b\\
\label{RMg}
&& \frac{I_H}{2}\leq b\leq I_H,
\end{eqnarray}
\end{subequations}
where $R_{u_1}$ and $ R_{u_2}$ are the achievable rates for decoding $s_1$ and $s_2$  at $U_1$, $U_2$ and defined at (\ref{Ru1}) and (\ref{Ru2}), respectively,  $R_{u_1\rightarrow D}$ and $R_{u_2\rightarrow D}$ are the achievable rates for decoding $s_D$ at  users $U_1$ and $U_2$ and defined at (\ref{RuD1}) and (\ref{RuD2}), respectively, $I_H$ is the maximum allowed input current to the VLC transmitter, and $\eta$ is the RF-to-VLC bandwidth ratio satisfying $\eta \in [0,1]$.  Constraint (\ref{RMb}) is the individual power constraint at the users (i.e., the transmit power must be less than or equal to the harvested power). On the other hand, constraint (\ref{RMc}) is imposed to achieve the required QoS constraint at the entrusted users, and  constraint (\ref{RMd}) is imposed to assure that the secrecy rate of the destination is not limited by the first hop (VLC link). Finally,  constraints (\ref{RMf}) and (\ref{RMg}) are imposed to guarantee that the input signal to the LED remains within the linear operational range to avoid clipping, and to guarantee the nonnegativity of the input signal. 

Note that problem (\ref{RM1}) is non-convex because the objective function is a difference between two concave functions. In what follows, we reformulate the problem in order to obtain a simple solution.
\subsection{Problem Reformulation}
By rewriting the objective function as
\begin{equation}
R_s=\frac{1}{2}\log_2\left(\frac{\sigma^2_{RF}+\mathbf{h}_D^H\mathbf{w}\mathbf{w}^H\mathbf{h}_D}{\sigma^2_{RF}+\mathbf{h}_E^H\mathbf{w}\mathbf{w}^H\mathbf{h}_E}\right),
\end{equation} 
problem (\ref{RM1}) can be equivalently written as 
\begin{subequations}
\label{RM2}
\begin{eqnarray}
&\displaystyle\max_{\mathbf{w}, b, P_1, P_2, P_D}&  \frac{\sigma^2_{RF}+\mathbf{h}_D^H\mathbf{w}\mathbf{w}^H\mathbf{h}_D}{\sigma^2_{RF}+\mathbf{h}_E^H\mathbf{w}\mathbf{w}^H\mathbf{h}_E},\\ 
&\text{s.t.}& \text{(\ref{RMb})-(\ref{RMg}).}
\end{eqnarray}
\end{subequations}

The objective function in (\ref{RM2}) is nonconvex. However, in the following, we provide an efficient solution that finds feasible users' powers, beamforming vector, and the DC-bias jointly, while achieving good performance. First, it is important to note that increasing the value of the variable $b$ increases the harvested energy, but decreases the total peak power $P_T=(\sqrt{P_1}+\sqrt{P_2}+\sqrt{P_D})^2$ that is used to transmit signals $s_1$, $s_2$, and $s_D$. In other words, increasing the DC-bias helps in increasing the objective function but tightens the QoS constraints (i.e., decreases $R_{u_i}$ and $R_{u_i\rightarrow D}$). Therefore, our goal of designing the DC-bias and the users' powers is to find the maximum value of $b$ that achieves the constraints or equivalently, to find the minimum values of $P_1$, $P_2$, and $P_D$ that achieve the QoS constraints. From  constraint (\ref{RMc}), when $i=1$, we can find the minimum value of $P_1$ that achieves constraint (\ref{RMc}) (when $i=1$) with equality and this value is the optimal solution of $P_1$. This is because increasing $P_1$ further leads to decreasing $b$ which consequently decreases $R_s$. Hence, the optimal value of $P_1$ is given by
\begin{equation}
\label{P1}
P_1^*= \frac{\sigma_v^2 (2^{2R_{th}}-1)}{c\nu^2\rho^2h_1^2}.
\end{equation}
Similarly, the optimal value of $P_2$ is the value that achieves constraint (\ref{RMc}) (when $i=2$) with equality. This is because increasing $P_2$ further leads to decreasing $b$ which consequently decreases $R_s$.  After finding $P_1^*$ using (\ref{P1}), the optimal value of $P_2$, using (\ref{RMc}) and when $i=2$,  is given by
\begin{equation}
\label{P2}
P_2^*=\frac{(2^{2R_{th}}-1)(\sigma_v^2+c\rho^2\nu^2h_2^2P_1^*)}{(c\rho^2\nu^2h_2^2)}.
\end{equation}
The challenge now is how to find the optimal $P_D$ since constraint (\ref{RMd}) is a function of $P_D$ and $R_s$. It can be seen that increasing $P_D$ leads to increasing the functions $R_{u_i\rightarrow D},\ i=1,2$ and to decreasing $R_s$. Therefore, the optimal $P_D$ is the minimum value of $P_D$ that achieves constraint (\ref{RMd}). Our approach to find feasible and good-performing $P_D$ and $\mathbf{w}$ is to first solve the problem under a given value of $P_D$, then use an outer loop to update $P_D$ using bisection search.

 From constraint (\ref{RMf}), if $P_D$ is given, the maximum DC-bias $b$ that achieves the constraints is given by
\begin{equation}
\label{bb}
b^*=I_H-\sqrt{P_1^*}-\sqrt{P_2^*}-\sqrt{P_D},
\end{equation}
and problem (\ref{RM1}) can be expressed as (if $P_D$ is given) 
\begin{subequations}
\label{RM3}
\begin{eqnarray}
&\displaystyle\max_{\mathbf{w}}&  \frac{\sigma^2_{RF}+\mathbf{h}_D^H\mathbf{w}\mathbf{w}^H\mathbf{h}_D}{\sigma^2_{RF}+\mathbf{h}_E^H\mathbf{w}\mathbf{w}^H\mathbf{h}_E}\\ 
&\text{s.t.}& \vert w_i \vert^2 \leq P_{r,i}(b), \ i=1,2.
\end{eqnarray}
\end{subequations}
In the following, we propose two approaches to tackle problem (\ref{RM3}).
\subsection{Charnes-Cooper with SDR Approach}  
In this approach, we apply Charnes-Cooper method and SDR to convert (\ref{RM3}) into a convex problem.
Defining $\mathbf{W}=\mathbf{w}^H\mathbf{w}$, $\mu=\frac{1}{\sigma^2_{RF}+\mathbf{h}_E^H\mathbf{w}\mathbf{w}^H\mathbf{h}_E}$, and $\mathbf{S}=\mu \mathbf{W}$, problem (\ref{RM3}) can be re-written as follows
\begin{subequations}
\label{RM4}
\begin{eqnarray}
&\displaystyle\max_{\mathbf{S}, \mu}&  \mu\sigma^2_{RF}+\text{tr}(\mathbf{S}\mathbf{H}_D)\\ 
&\text{s.t.}& \mu\sigma^2_{RF}+\text{tr}(\mathbf{S}\mathbf{H}_E)=1,\\
&& \text{tr}(\mathbf{S}\mathbf{E}_i) \leq \mu P_{r,i}(b), \ i=1,2,\\
&& \mathbf{S} \succcurlyeq 0,\ \ \text{Rank}(S)=1,
\label{Srank}
\end{eqnarray}
\end{subequations}
where $\mathbf{H}_D=\mathbf{h}_D\mathbf{h}_D^H$, $\mathbf{H}_E=\mathbf{h}_E\mathbf{h}_E^H$, $\mathbf{E}_i$ is a $2\times 2$ matrix with its $i$th diagonal entry equals to one and all the other entries equal to zero, and tr$(.)$ is the trace function. Constraint (\ref{Srank}) is equivalent to  $\mathbf{W}=\mathbf{w}^H\mathbf{w}$. Problem (\ref{RM4}) is not convex because of the rank constraint. However, if we relax the rank constraint, the problem becomes convex and can be solved efficiently using the interior point method \cite{Boyd}, which can be implemented using CVX \cite{cvx}.  Since the rank constraint is relaxed, the resulting $\mathbf{S}$ matrix is not guaranteed to achieve the optimal solution.  The authors of \cite{huang2009rank} showed that if the number of trace constraints less than or equal three, the resulting $\mathbf{S}$ from (\ref{RM4}) is of rank one. However, in some cases, CVX does not produce an absolute rank one $\mathbf{S}$ (i.e., in some cases CVX produce a matrix $\mathbf{S}$, where the second maximum eigenvalue is close to zero but not exactly zero). This means that, theoretically, the proposed SDR approach provides the optimal solution, but this optimal solution is not numerically guaranteed. 
It was shown that the computational complexity of SDR approach is in the order of $O(N^7)$, where $N$ is the length of $\mathbf{w}$ \cite{sidiropoulos2006transmit,obeed2018efficient}.
\subsection{Null Space Beamforming (Zero Forcing (ZF)) Approach}
In this section, we propose a simpler approach to find a solution for (\ref{RM3}). We propose to design the beamforming vector to null the transmitted signal at the eavesdropper. This approach performs well  when the eavesdropper's channel is much better than the destination's channel. 

We set $w_1=ah_{E,2}^H$ and $w_2=-ah_{E,1}^H$, where $a$ is a scalar value that should be selected to maximize the secrecy capacity while satisfying the constraints. With this design for the vector $\mathbf{w}$, the optimization problem (\ref{RM3}) can be expressed as follows

\begin{subequations}
\label{E3}
\begin{eqnarray}
&\displaystyle\max_{a}&  a(h_{E,2}^Hh_{D,1}-h_{E,1}^Hh_{D,2})\\ 
&\text{s.t.}& a^2\vert h_{E,2}\vert^2 \leq P_{r,1}(b), \\
&& a^2\vert h_{E,1}\vert^2 \leq P_{r,2}(b).
\end{eqnarray}
\end{subequations}
The optimal value of $a$ in (\ref{E3}) can be shown to be given by \begin{equation}
a^*=\min\left(\frac{\sqrt{P_{r,1}(b)}}{\vert h_{E,2}\vert},\frac{\sqrt{P_{r,2}(b)}}{\vert h_{E,1}\vert}\right).
\end{equation} 
It is important to note that the computational complexity of the ZF approach is much lower than that of the SDR approach.
\subsection{Joint Destination's Power and Beamforming Solution}
 In this section, we provide the overall algorithm that solves problem (\ref{RM1}). Earlier, we showed that the optimal $P_1$ and $P_2$ are given by (\ref{P1}) and (\ref{P2}), respectively. We also showed that the optimal $P_D$ is the minimum value of $P_D$ that achieves constraint (\ref{RMd}). Since the minimum value of DC-bias $b$ is not less than $\frac{I_H}{2}$ (constraint (\ref{RMf})) and because of constraint (\ref{RMd}), the optimal $P_D$ is bounded by  
\begin{equation}
\label{PDb}
G(R_s^{(0)})\leq P_D^* \leq
\big(\frac{I_H}{2}-\sqrt{P_1^*}-\sqrt{P_2^*}\big)^2,
\end{equation} 
where 
\begin{multline*}
\Scale[0.99]{G(R_s^{(0)})=\max\bigg(\frac{(2^{\eta R_{s}^{(0)}}-1)(\sigma_v^2+c\rho^2\nu^2h_1^2P_1^*+c\rho\nu^2h_1^2P_2)}{(c\rho^2\nu^2h_1^2)}},\\ \Scale[0.99]{\frac{(2^{\eta R_{s}^{(0)}}-1)(\sigma_v^2+c\rho^2\nu^2h_2^2P_1^*+c\rho\nu^2h_2^2P_2^*)}{(c\rho^2\nu^2h_2^2)}\bigg)},
\end{multline*}
and $R_s^{(0)}$ is the minimum secrecy rate that resulted by setting $P_D=\big(\frac{I_H}{2}-\sqrt{P_1^*}-\sqrt{P_2^*}\big)^2$. It can be seen that there is only a unique value of $P_D$ within the range (\ref{PDb}) that maximizes $R_s$ and achieves the constraints. This is because as we decrease $P_D$, the value of $R_s$ increases and the value of $\min(R_{u_1\rightarrow D},R_{u_2\rightarrow D})$ decreases, which means that there is only one value that maximizes $R_s$ and achieves constraint (\ref{RMd}) with equality. To find the optimal $P_D^*$, we can apply the bisection method, where in each step, we have to solve the problem for $\mathbf{w}$, using either the proposed Charnes-cooper with SDR or the null space beamforming.
\begin{algorithm}
\SetAlgoLined
 \label{algor1}
 Find $P_1$ and $P_2$, using (\ref{P1}) and (\ref{P2}), respectively\;
 Set $a_1=G(R_s^{(0)})$ and $a_2=\big(\frac{I_H}{2}-\sqrt{P_1^*}-\sqrt{P_2^*}\big)^2$\;
 \For {for $i=1:M$}{
 Set $P_D^{(i)}=\frac{a_1+a_2}{2}$, and find $\min(R_{u_1\rightarrow D}^{(i)},R_{u_2\rightarrow D}^{(i)})$\;
 Find $\mathbf{w}^{(i)}$ and then $R_s^{(i)}$ using either SDR or ZF approach\;
 \eIf {$ R_s^{(i)}-\min(R_{u_1\rightarrow D}^{(i)},R_{u_2\rightarrow D}^{(i)})< 0$}{
 Set $a_2=P_D^{(i)}$\;}
 {
 Set $a_1=P_D^{(i)}$\;
 }
 \If {$\vert n-m\vert\leq \epsilon$}
  {Break;}
 }
 Find $b^*=I_H-\sqrt{P_1^*}-\sqrt{P_2^*}-\sqrt{P_D^*}$;
 \caption{Find joint $P_1$, $P_2$, $b$, $P_D$, and $\mathbf{w}$ solution.}
 \end{algorithm}

 where $M$ is the maximum number of iterations and $\epsilon$ is a positive value selected to be very small to guarantee the convergence.
\section{Performance Optimization without Eavesdropper's CSI}

In this section, we assume that the entrusted users and the VLC AP do not know the instantaneous CSI information of the eavesdropper, but the statistical information of the eavesdropper is known. In this case, we propose three solutions to improve the secrecy rate. The first is a baseline solution based on the MRT approach, where the beamforming vector is designed to maximize the received data rate at the destination. The others two approaches are based on generating artificial noise to confuse the possible eavesdropper. In all approaches, we allocate the first hop parameters to achieve the constraints and achieve the required QoS at the entrusted users.
\subsection{Beamforming Approach (Baseline Approach)}

In this approach, we assume that the entrusted users assign all their power to transmit the destination's message. Therefore, the achievable secrecy rate is given by 
\begin{equation}
\bar{R}_s=\min\big(R_{u_1\rightarrow D}, R_{u_2\rightarrow D},R_{s,RF}),
\end{equation}
where $R_{s,RF}$ is the average secrecy rate of the RF hop and it is given by  $$R_{s,RF}=E\bigg[\frac{1}{2}\log_2\left(\frac{\sigma^2_{RF}+\mathbf{h}_D^H\mathbf{w}\mathbf{w}^H\mathbf{h}_D}{\sigma^2_{RF}+\mathbf{h}_E^H\mathbf{w}\mathbf{w}^H\mathbf{h}_E}\right)\bigg],$$ where $E[.]$ is the expectation function \cite{Li2011,Chaaban2016}.
The problem can then be expressed as follows
\begin{subequations}
\label{RU1}
\begin{eqnarray}
&\displaystyle\max_{\mathbf{w}, b, P_1, P_2, P_D}&  \bar{R}_s\\
\label{RUb}
&\text{s.t.}&  \vert w_i \vert^2 \leq P_{r,i}(b), \ i=1,2\\ 
\label{RUc}
&&  R_{u_i}\geq R_{th},\ i=1,2 ,\\
\label{RUd}
&& \sqrt{P_1}+\sqrt{P_2}+\sqrt{P_D} \leq I_H-b\\
\label{RUe}
&& \frac{I_H}{2}\leq b\leq I_H.
\end{eqnarray}
\end{subequations}
Solving problem above is not straightforward because the objective function is not concave and because of the expectation term. However, we propose a simple, yet efficient, solution and adopt this solution as a baseline to the second approach. From (\ref{RU1}), it can be seen that functions $R_{u_1\rightarrow D},\ R_{u_2\rightarrow D},\ R_{u_1}, $ and $ R_{u_2}$ do not rely on the variable $\mathbf{w}$, while $R_{s,RF}$ relies on all variables. 
Hence, we propose to allocate $P_1,\ P_2,\ P_D$ and $b$ to maximize the functions $R_{u_2\rightarrow D},\ R_{u_2\rightarrow D}$ and achieve the constraints in (\ref{RUc})-(\ref{RUe}), while $\mathbf{w}$ is selected to maximize the function $R_{s,RF}$. To do so, the variables $b, P_1, $ and $ P_2$ must be at their minimum values and $P_D$ must be at its highest value. Therefore, $b, P_1, $ and $ P_2$ are given by  $b=\frac{I_H}{2}$, $P_1=\frac{\sigma_v^2 (2^{2R_{th}}-1)}{c\nu^2\rho^2h_1^2}$, and $P_2=\frac{(2^{2R_{th}}-1)(\sigma_v^2+c\rho^2\nu^2h_2^2P_1)}{(c\rho^2\nu^2h_2^2)}$, while $P_D=(\frac{I_H}{2}-\sqrt{P_1}-\sqrt{P_2})^2$. For the given $b$, $P_1$, $P_2$, $P_D$, problem (\ref{RU1}) boils down to
\begin{subequations}
\label{RU3}
\begin{eqnarray}
&\displaystyle\max_{\mathbf{w}}&  R_{s,RF}\\
\label{RU3b}
&\text{s.t.}&  \vert w_i \vert^2 \leq P_{r,i}(b), \ i=1,2.
\end{eqnarray}
\end{subequations}
The entrusted users transmit the signal so as to maximize the SNR of  the destination. Hence, the solution is given by $w_i=\frac{\sqrt{P_{r,i}(b)}}{\vert h_{D,i}\vert}h_{D,i},\ i=1,2$. Note that this beamforming approach is not a powerful solution for secrecy improving the secrecy rate if the eavesdropper's CSI is not known, because it only aims to maximize the SNR of the destination, while ignoring the presence of the eavesdropper. A better solution is to divide the power between beamforming the signal to the destination and sending a jamming signal to confuse the eavesdropper.  

\subsection{Artificial Noise with SDR Approach}
In this approach, the entrusted users divide their harvested power into two portions: one for forwarding the destination's message, and the other jamming. Since the destination's channels information is available at the entrusted user, the jamming signal can be designed to be orthogonal to the legitimate destination's channel. Defining $\vert n_{a,i}\vert^2$ as the power assigned for the jamming signal at user $i$, the transmitted signal of user $i$ is $\bar{y}_i=w_is_D+n_{a,i}z,$ where $s_D$ and $z$ are the destination message and the noise signal with $E[\vert s_D \vert^2]=1$ and $E[\vert z \vert^2]=1$, respectively. The jamming signal at the destination can be nulled if we set that $n_{a,1}=\beta h_{D,2}$ and $n_{a,2}=-\beta h_{D,1}$, where $\beta$ is a scalar that can be selected to maximize the power of the jamming signal.
Therefore, the average achievable secrecy rate is given by 
\begin{equation}
\tilde{R}_s=\min\big(R_{u_1\rightarrow D}, R_{u_2\rightarrow D},\tilde{R}_{s,RF}),
\end{equation}
where \begin{multline*}
\tilde{R}_{s,RF}=E\bigg[\frac{1}{2}\log_2\left(1+\frac{\mathbf{h}_D^H\mathbf{w}\mathbf{w}^H\mathbf{h}_D}{\sigma^2_{RF}}\right)-\\
\frac{1}{2}\log_2\left(1+\frac{\mathbf{h}_E^H\mathbf{w}\mathbf{w}^H\mathbf{h}_E}{\sigma^2_{RF}+\mathbf{n}_a^H\mathbf{h}_E\mathbf{h}_E^H\mathbf{n}_a}\right)\bigg].
\end{multline*}

The total transmit power at the entrusted users is  $\Vert \mathbf{w}\Vert^2+\Vert \mathbf{n}_a\Vert^2$. It can be seen that devoting more power for the jamming signal would confuse the eavesdropper more, but this decreases the received signal power at the destination. Therefore, we first set the minimum required QoS at the destination and then allocate the remaining power to minimize the average achievable rate at the eavesdropper. To do so, we formulate the problem as maximizing the artificial noise power subject to achieving the required QoS at the destination. The problem can then be formulated as follows
\begin{subequations}
\label{RM5}
\begin{eqnarray}
&\displaystyle\max_{\mathbf{w},\mathbf{n}_a, b, P_1, P_2, P_D}&  \Vert \mathbf{n}_a\Vert^2\\
\label{RM5b}
&\text{s.t.}&\Scale[0.9]{\vert w_i \vert^2+\vert n_{a,i} \vert^2 \leq P_{r,i}(b),\ i=1,2,}\\
 \label{RM5c}
&&  \tilde{R}_D\geq R_{th,D}\\
\label{RM5d}
&&  R_{u_i}\geq R_{th},\ i=1,2 ,\\
\label{RM5e}
&& R_{u_i\rightarrow D}\geq \eta R_{th,D},\ i=1,2,\\
\label{RM5g}
&& \Scale[0.97]{\sqrt{P_1}+\sqrt{P_2}+\sqrt{P_D} \leq I_H-b}\\
\label{RM5h}
&& \frac{I_H}{2}\leq b\leq I_H,
\end{eqnarray}
\end{subequations}
where $\tilde{R}_D=\frac{1}{2}\log_2\left(1+\frac{\mathbf{h}_D^H\mathbf{w}\mathbf{w}^H\mathbf{h}_D}{\sigma^2_{RF}}\right)$ and $R_{th,D}$ are the achievable rate and the minimum required data rate at the destination, respectively.  Constraints in (\ref{RM5e}) are imposed to make sure that the average secrecy rate is not limited with the first hop (i.e., the VLC hop).

First, we should note that the optimal values of the messages' powers of the entrusted users can be derived similar to what is conducted in Section \ref{CSI}, where equations  (\ref{P1}) and (\ref{P2})  can be, respectively, used to find  $P_1$ and $P_2$. In problem (\ref{RM5}), it can be shown that the optimal $P_D$ is the minimum value that achieves constraint (\ref{RM5d}). Hence, the optimal $P_D$ is given by 
\begin{multline}
P_D^*=\Scale[1.1]{\max\bigg(\frac{(2^{\eta R_{th,D}}-1)(\sigma_v^2+c\rho^2\nu^2h_1^2P_1^*+c\rho\nu^2h_1^2P_2)}{(c\rho^2\nu^2h_1^2)}},\\ \Scale[1.1]{\frac{(2^{\eta R_{th,D}}-1)(\sigma_v^2+c\rho^2\nu^2h_2^2P_1^*+c\rho\nu^2h_2^2P_2^*)}{(c\rho^2\nu^2h_2^2)}\bigg)}.
\end{multline}
Therefore, the minimum value of the DC-bias $b$ that can achieve the constraint is given by 
\begin{equation}
\label{bb2}
b^*=I_H-\sqrt{P_1^*}-\sqrt{P_2^*}-\sqrt{P_D^*}.
\end{equation}


For the given $P_1$, $P_2$, $P_D$, and $b$, solving problem (\ref{RM5}) (in terms only of $\mathbf{w},\mathbf{n}_a$) is not straightforward since the norm function is convex (not concave). Since $P_1$, $P_2$, $P_D$, and $b$ are given and $n_{a,1}=\beta h_{D,2}$, $n_{a,2}=-\beta h_{D,1}$, problem (\ref{RM5}), can be equivalently written as  
\begin{subequations}
\label{RM7}
\begin{eqnarray}
&\displaystyle\max_{\mathbf{w},\mathbf{n}_a}&  \beta\\
\label{RM7b}
&\text{s.t.}&\vert w_i \vert^2+\beta^2\vert h_{D,1} \vert^2 \leq P_{r,i}(b),\ i=1,2,\\
 \label{RM7c}
&&  \mathbf{h}_D^H\mathbf{w}\mathbf{w}^H\mathbf{h}_D \geq \sigma^2_{RF}(2^{2R_{th,D}}-1).
\end{eqnarray}
\end{subequations}
To simplify problem (\ref{RM7}) which is not convex, we introduce  a variable $\mathbf{W}=\mathbf{w}\mathbf{w}^H$ and use the SDP approach with relaxation. Thus (\ref{RM7}) can be re-expressed as follows
\begin{subequations}
\label{RM8}
\begin{eqnarray}
&\displaystyle\max_{\mathbf{w},\mathbf{n}_a}&  \beta\\
\label{RM8b}
&\text{s.t.}& \text{tr}(\mathbf{W}\mathbf{E_i})+\beta^2\vert h_{D,1} \vert^2 \leq P_{r,i},\ i=1,2,\\
 \label{RM8c}
&&  \text{tr}(\mathbf{W}\mathbf{H_D}) \geq \sigma^2_{RF}(2^{2R_{th,D}}-1),\\
\label{RM8d}
&& \mathbf{W} \succcurlyeq 0,\ \  \text{Rank}(\mathbf{W})=1,
\end{eqnarray}
\end{subequations}  
where $\mathbf{E}_i$ is $2\times 2$ matrix with all its entries equal to zero except the $i$th diagonal entry which is equal to one. Constraint (\ref{RM8d}) guarantees that $\mathbf{W}=\mathbf{w}\mathbf{w}^H$. To simplify problem (\ref{RM8}) further, we drop the rank constraint so that it can be solved efficiently using the interior point method \cite{Boyd}, and  can be implemented using CVX \cite{cvx}. The resulting $\mathbf{W}$ from (\ref{RM8}) is of rank one (because the number of trace constraints is not larger than three \cite{huang2009rank}). Nevertheless, the CVX may not produce a rank one $\mathbf{W}$ matrix (i.e., the second highest eigenvalue of the resulting matrix is close to zero, but not zero), in which case a randomization method is used to find a good solution using the resulting $\mathbf{W}$ \cite{sidiropoulos2006transmit}. 

\subsubsection{Artificial Noise with Maximum Ratio Transmission}
Since the complexity of using SDR approach is high, we propose a simpler solution for (\ref{RM7}) by using  MRT instead of SDR.  We select $\mathbf{w}$ to be aligned with $\mathbf{h}_D$ to maximize the expression $\mathbf{h}_D^H\mathbf{w}\mathbf{w}^H\mathbf{h}_D$. Therefore, we select $w_1=\alpha_1h_{D,1}$ and $w_2=\alpha_2h_{D,2}$, where $\alpha_1$ and $\alpha_2$ are scalar values selected to achieve the constraints. Using this substitution, problem (\ref{RM7}) can be rewritten as follows

\begin{subequations}
\label{RM9}
\begin{eqnarray}
&\displaystyle\max_{\alpha_1,\alpha_2,\beta}&  \beta\\
\label{RM9b}
&\text{s.t.}&\alpha_1^2\vert h_{D,1} \vert^2+\beta^2\vert h_{D,2} \vert^2 \leq P_{r,1},\\
\label{RM9c}
&&\alpha_2^2\vert h_{D,2} \vert^2+\beta^2\vert h_{D,1} \vert^2 \leq P_{r,2},\\
 \label{RM9d}
&&\Scale[0.87]{\alpha_1 \vert h_{D,1}\vert^2+\alpha_2\vert h_{D,2}\vert^2 \geq \sigma_{RF}\sqrt{2^{2R_{th,D}}-1}},\\
&& \beta\geq 0.
\end{eqnarray}
\end{subequations}
Problem (\ref{RM9}) is convex and can be solved efficiently using the CVX. In the following section, we show some simulation results to assess the secrecy performance of the proposed approaches with changing some of the system's parameters.

\begin{table}[!t]
\centering
\caption{Simulation Parameters}
\label{table1}
\begin{tabular}{|p{.25\textwidth} | p{.17\textwidth} |}
\hline
  Parameter's Name& Parameter's Value\\
 \hline

 VLC AP  Bandwidth, $B$ & $20$ MHz  \\

  The physical area of PDs, $A_{p}$ & $1$\ cm$^2$ \\
   Half-intensity radiation angle, $\theta_{1/2}$ & $60^o$\  \\
  Gain of optical filter, $g_{of}$ & $1$  \\
   Optical-to-electrical conversion factor, $\rho$& $0.53$ [A/W]\\
   Electric-to-optical conversion factor, $\nu$ & 10 W/A\\
   Refractive index, $n$ & 1.5 \\
   Noise PSD of LiFi, $N_0$ & $10^{-21}$\ A$^2$/Hz  \\
  Maximum input bias current, $I_H$ & $600$ mA  \\

  Minimum input bias current, $I_L$ & $0$ mA  \\
  Fill factor, $f$ &0.75\\
  
  Thermal voltage, $V_t$ & 25 mV \\

  Dark saturation current of the PD, $I_0$ & $10^{-10}$ A\\
  LED height, &$3$ m\\
  User height & $0.85$\\
  \hline

  RF    \\
  \hline
   Bandwidth & 16 MHz\\
   PSD of the noise & -174 dBm/Hz\\
  The  breakpoint distance & 5 m\\
  Angle of arrival/departure of LoS & 45$^o$\\
  Central carrier frequency  & 2.4 GHz\\
  Shadow fading standard deviation (after the breakpoint) & 5 dB  \\
  Shadow fading standard deviation (before the breakpoint) & 3 dB  \\
  \hline
\end{tabular}

 \end{table}
  
\section{Simulation Results}
We evaluate the proposed solutions that are used to allocate the users'  powers, the DC-bias, the beamforming vector, and the jamming vector to improve the secrecy rate of the Co-NOMA hybrid VLC/RF system. We examine the effect of the quality of the channels of the destination and the eavesdropper by changing their distances from the VLC cell center (circular coverage area). We also examine the effect of the required data rates of the entrusted users and of the destination in the case of unknown eavesdropper's CSI. Simulation parameters are provided in Table \ref{table1}. The entrusted users are randomly distributed in a circle of radius $2$ m around the cell center at (0,0) on the floor level.  Monte-Carlo simulation is used to assess the proposed solutions. Each point in the figures is the result of 300 different users' positions. We evaluate the secrecy performance for different distances between the destination and the eavesdropper on the one hand, and the entrusted circle center on the other hand ($D_D$ and $D_E$), and for different required QoS for the users ($R_{th}$ and $R_{th,D}$).   

\subsection{The CSI of the Eavesdropper is Known}
\begin{figure}[!t]
\centering
\includegraphics[width=3.7in]{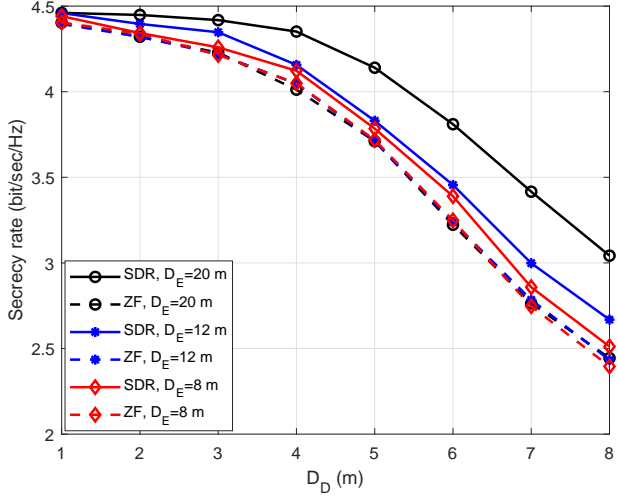}
\caption{Secrecy rate versus the distance between the legitimate destination and the center of the covered circle when $R_{th}=2$.}
\label{DD_SR}
\end{figure}

\begin{figure}[!t]
\centering
\includegraphics[width=3.7in]{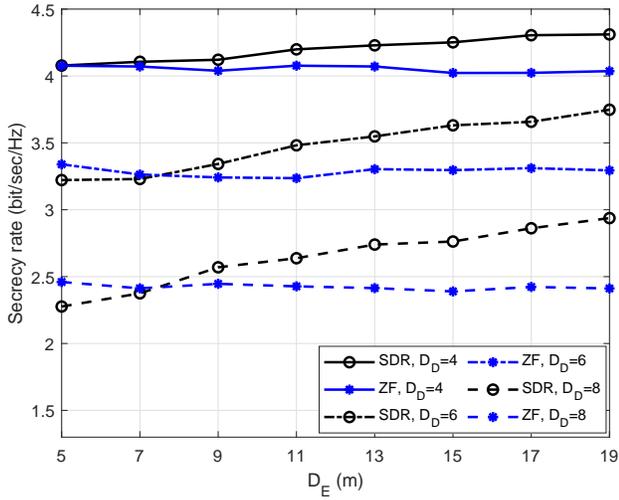}
\caption{Secrecy rate versus the distance between the eavesdropper and the center of the covered circle when $R_{th}=2$ and the CSI of the eavesdropper is known.}
\label{DE_SR}
\end{figure}

(n Fig. \ref{DD_SR}, we show the effect of increasing the distance of the legitimate destination from the cell center with different eavesdropper's distance values. As expected, the secrecy rate decreases as the distance between the destination and the entrusted users increases and as the distance between the eavesdropper and the entrusted users decreases.  Fig. \ref{DD_SR} also shows that the proposed SDR approach with Charnes-Cooper outperforms the proposed ZF approach. The performance of SDR approach is improved by decreasing the eavesdropper's channel quality, while the null space beamforming approach does not get the benefit of decreasing the eavesdropper's channel quality. To make this point clearer we show the effect of degrading the eavesdropper's channel on the secrecy capacity in Fig. \ref{DE_SR}.

Fig. \ref{DE_SR} shows that the secrecy rate of the null space beamforming approach approximately stays fixed as we increase the distance of the eavesdropper. In contrast, in the SDR approach, the secrecy rate increases as the channel of the eavesdropper deteriorates. Also, it can be seen that the null space beamforming approach performs better than the SDR when the channel of the eavesdropper is much better than the channel of the legitimate destination. Fig. \ref{DE_SR} shows that as the channel of the eavesdropper gets much closer to the entrusted users, it is better to apply the null space beamforming approach than to apply the SDR approach.  The relaxed rank constraint in SDR approach affects its optimality, and this is why SDR cannot perform better than the null space beamforming when the eavesdropper is so close to the transmitters. However, the SDR approach performs better than null space beamforming when the channel quality of the eavesdropper is less than or closer to the channel quality of the destination.  

\begin{figure}[!t]
\centering
\includegraphics[width=3.7in]{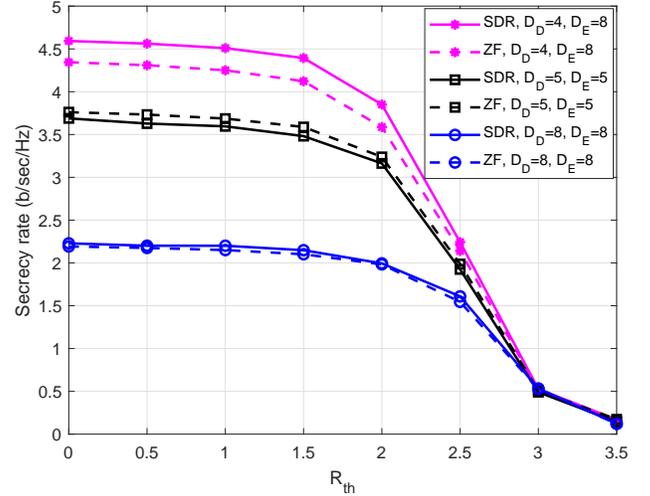}
\caption{Secrecy rate versus the minimum required $R_{th}$ at the entrusted users when the eavesdropper's CSI is known.}
\label{Rth_SR}
\end{figure}

In Fig. \ref{Rth_SR}, we show how the minimum required QoS at the entrusted users affects the secrecy performance.  The figure shows the effect of the required $R_{th}$ with different destination and eavesdropper channel qualities. Increasing $R_{th}$ leads to decreasing the harvested power at the entrusted users and to decreasing the achievable rate of the destination coming from the VLC link (i.e., $R_{u_i\rightarrow D}\ \forall i$). Since we assumed that the VLC bandwidth is higher than that of the RF, the effect of decreasing  $R_{u_i\rightarrow D}$ on the secrecy performance does not appear at the smaller values of $R_{th}$ (only decreasing the harvested power is affecting the secrecy rate). However, with increasing $R_{th}$ up to some point, the secrecy performance starts to be determined by the first hop (the VLC link). This is the reason why the effect of $R_{th}$ is significant at higher values of $R_{th}$.  
\subsection{The CSI of the Eavesdropper is Unknown} 

\begin{figure}[!t]
\centering
\includegraphics[width=3.7in]{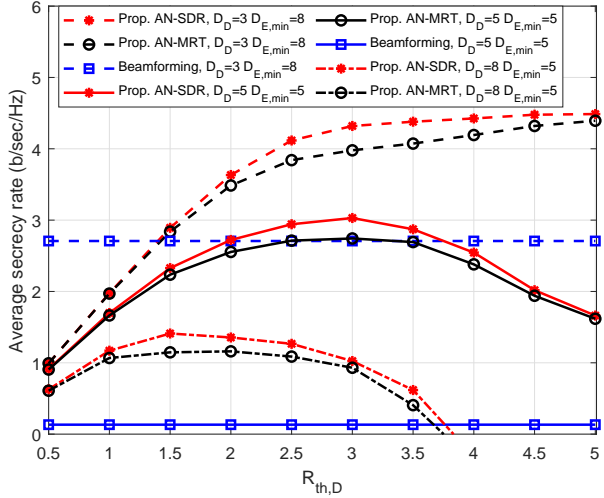}
\caption{Secrecy rate versus the minimum required $R_{thD}$ at the legitimate destination when the eavesdropper's CSI is unknown.}
\label{RthD_SR}
\end{figure}

 Fig. \ref{RthD_SR} shows how the minimum required data rate at the legitimate destination affects the secrecy performance. It shows that the amount of the required artificial noise to maximize the secrecy capacity depends on the channel quality of both the eavesdropper and the destination. Specifically, selecting the value of $R_{thD}$ depends on the location of the eavesdropper. If the eavesdropper is much closer to the entrusted users than the destination, minimizing the $R_{thD}$ to maximize the artificial noise power is the appropriate strategy to improve the secrecy performance. On the other hand, it is not wise to put a high power on emitting the jamming signal if the eavesdropper is much farther from the entrusted users than the destination. The figure also shows that the beamforming approach is not a function of $R_{th,D}$, because this approach beamforms all the available power to the direction of the legitimate destination. It is shown that the proposed artificial noise approaches significantly outperform the beamforming approach  in all cases except that the required $R_{th,D}$ at the destination is small and the destination is much closer to the transmitters than the the eavesdropper.

\begin{figure}[!t]
\centering
\includegraphics[width=3.7in]{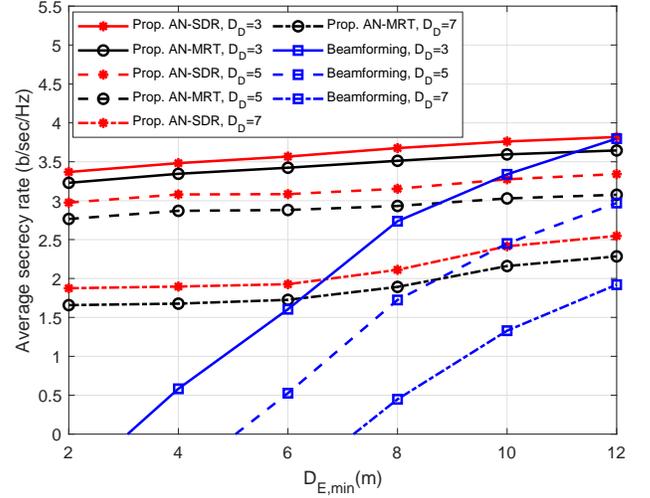}
\caption{Secrecy rate versus the distance between the eavesdropper and the center of the covered circle when $R_{th}=2$ and the CSI of the eavesdropper is unknown, $R_{th,D}=2$ and $D_D=5$ m.}
\label{DE_SR2}
\end{figure}

Fig. \ref{DE_SR2} shows how changing the eavesdropper's and the destination's channel quality would affect the secrecy rate. It can be seen that decreasing the eavesdropper's channel (increasing $D_{E,min}$, where $D_{E,min}$ is minimum eavesdropper's distance that the system can achieve such secrecy rate) slightly improves the secrecy performance when the artificial noise is applied, while decreasing the destination's channel quality significantly decreases the average secrecy rate. In contrast, the effect of decreasing the eavesdropper's channel quality is high when the beamforming approach is applied. The small effect of the eavesdropper's channel, when the artificial noise is applied, is due to the fact that increasing the eavesdropper's distance decreases both the received signal power and noise power.  However, the proposed artificial noise approaches outperform the beamforming approach except in the case where the eavesdropper is so far from the transmitters. In addition, if the eavesdropper is located closer to the transmitters than the destination, the beamforming approach cannot provide a non-zero secrecy rate.


\begin{figure}[!t]
\centering
\includegraphics[width=3.7in]{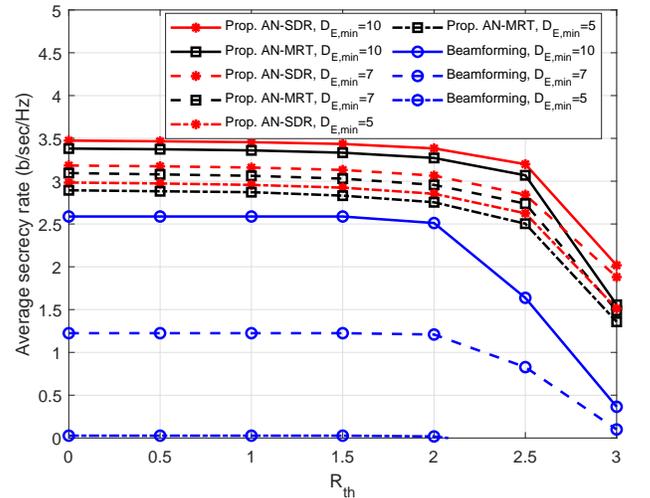}
\caption{Secrecy rate versus the minimum required $R_{th}$ at the entrusted users when the eavesdropper's CSI is unknown.}
\label{Rth_SR2}
\end{figure}

Fig. \ref{Rth_SR2} shows the effect of the minimum required of data rates at the entrusted users on the secrecy rate. Clearly, increasing $R_{th}$ leads to increasing the minimum required powers for transmitting the messages at the first hop, which leads to decreasing $P_D$ and the harvested power that is used for beamforming and jamming in the second hop. In addition, increasing the required power at the entrusted users decreases the available power to be assigned for the destination message in the first hop, which makes the average secrecy rate limited by the first hop. In other words, the required QoS at the entrusted users obviously affects the secrecy performance at the legitimate destination. 


Figures \ref{RthD_SR}, \ref{DE_SR2}, and \ref{Rth_SR2} show that the artificial noise SDR approach outperforms the artificial noise with MRT approach with the different values of $R_{th,D}$, $D_D$, $D_{E,min}$, and $R_{th}$. This is because the fact that in the SDR approach, we optimize the beamforing vector $\mathbf{w}$ and the power of the artificial noise, while in Approach 2, we just focus on optimizing the powers of the beamforming and the jamming signals under the assumption that $\mathbf{w}$ is designed to match the destination channel, and $\mathbf{n}_a$ is designed to cancel the jamming signal at the destination. However, the artificial noise with MRT approach is simpler than the SDR approach and generally provides a much better performance than the baseline (beamforming) approach. The figures also show that the beamforming approach is unable to provide a positive average secrecy rate if the eavesdropper is closer to the entrusted users than the destination.


\section{Conclusion}

This paper evaluated and optimized the physical layer security in Co-NOMA hybrid VLC/RF system. With the system model that consists of a single VLC AP, two entrusted users, one legitimate destination, and one eavesdropper, we considered the problem of maximizing the secrecy rate at the destination, under QoS constraints, by allocating the messages' powers, DC-bias, and the beamforming vector. Under the assumption that the eavesdropper's CSI is known, we considered the problem of maximizing the secrecy capacity subject to QoS constraints. We proposed two solutions for such non-convex optimization problem: one is by using Charnes-Cooper and SDR and the other is by designing the beamforming vector to eliminate the signal at the eavesdropper. Simulation results showed that when the eavesdropper's CSI is known, the proposed SDR with  Charnes-Cooper method performs better than the zero-forcing approach if the eavesdropper is a little bit far from the center of the area of the entrusted users, while if the eavesdropper gets much closer to the transmitters, it is better to null the transmitted signal at the eavesdropper. When the CSI of the eavesdropper is assumed to be unknown, we considered three approaches: beamforming, artificial noise with SDR, and artificial noise with MRT. Numerical results showed that the artificial noise with SDR slightly outperforms the artificial noise with MRT and both artificial noise based approaches significantly outperform the beamforming approach in terms of average secrecy rate. Numerical results also showed that whether the CSI is available or not, the required QoS at the entrusted users significantly compromises the secrecy performance.


%

\bibliography{mylib}
\bibliographystyle{IEEEtran}

\end{document}